\documentclass{PoS}
\usepackage{amsmath}
\usepackage{adjustbox}
\usepackage{caption}
\usepackage{subcaption}
\usepackage{array}
\usepackage{url}
\usepackage{booktabs}
\usepackage{slashed}
\usepackage{color}
\usepackage{tikz}
\usetikzlibrary{shapes}
\newcommand{\invtriangle}{\raisebox{0pt}{\tikz{\node[draw,scale=0.3,regular polygon, regular polygon sides=3,fill=black!45!green,rotate=180](){};}}}

\newcommand{\markercircle}{\raisebox{0.5pt}{\tikz{\node[draw,scale=0.4,circle,fill=black!20!blue](){};}}}

\newcommand{\markerdiamond}{\raisebox{0pt}{\tikz{\node[draw,scale=0.4,diamond,fill=black!10!gray](){};}}}

\title{Proton decay matrix element on the lattice with physical pion mass}

\ShortTitle{Proton Decay Matrix element}

\author{\speaker{Jun-Sik Yoo }${}^{ a}$, Yasumichi Aoki${}^{b,c}$\footnote{Present Address: KEK and RIKEN Center for Computational Science}, Taku Izubuchi${}^{b,d}$, Sergey Syritsyn ${}^{a,b}$ \\
  \llap{${}^a$} Department of Physics and Astronomy, Stony Brook University,
      Stony Brook, NY 11794, USA \\
  \llap{${}^b$} RIKEN/BNL Research Center, Brookhaven National Laboratory,
      Upton, NY 11973, USA\\
  \llap{${}^c$} High Energy Accelerator Research Organization (KEK), Tsukuba, Ibaraki 305-0801, Japan\\
  \llap{${}^d$} Physics Department, Brookhaven National Laboratory,
      Upton, NY 11973, USA\\
        E-mail: \email{jun-sik.yoo@stonybrook.edu}}

\abstract{Proton decay is one of possible signatures of baryon number violation, which has to exist to explain the baryon asymmetry and the existence of nuclear matter. Proton decays must be mediated through effective low-energy baryon number violating operators made of three quarks and a lepton. We calculate matrix elements of these operators between the proton and various meson final states using the direct method. We report on preliminary results of matrix element calculation done with the 2+1 dynamical flavor domain wall fermions at the physical point for the first time.
}

\FullConference{The 36th Annual International Symposium on Lattice Field Theory - LATTICE2018\\
		22-28 July, 2018\\
		Michigan State University, East Lansing, Michigan, USA.}

\begin{document}
\section{Introduction}
\label{intro}
Baryon asymmetry is the imbalance between matter and anti-matter in the universe. Matter anti-matter asymmetry may be generated only if three necessary conditions are met \cite{Sakharov:1967dj}. Proton decay would be evidence for the baryon number violation, which is one of the conditions. The standard model lagrangian does not violate the baryon number, so we have to consider effective operators originating from beyond the standard model. Grand Unified theory(GUT) and Supersymmetric Grand Unified theory(SUSY-GUT) hypothesize a larger gauge group unifying the standard model interactions. The larger gauge group contains new interactions between leptons and quarks that can lead to proton decay. The large energy scale of (SUSY-) GUTs $\Lambda \sim 10^{16}$ GeV allows us to use effective operators for such processes with baryon number violation.

Evaluation of the matrix elements of these effective operators was attempted in numerous ways including non-relativistic quark model\cite{Gavela:1981cf}, chiral lagrangian \cite{Kaymakcalan:1983uc}, and bag model \cite{Okazaki:1982eh,Martin:2011nd}. Model-independent lattice QCD calculations of proton decay matrix elements have also been done \cite{Aoki:1999tw,Aoki:2006ib,Aoki:2008ku,Aoki:2013yxa,Aoki:2017puj}. However, previous lattice calculations have been performed with heavy pion masses, and the chiral extrapolation to the physical mass introduces systematic uncertainties. It was also suggested that the proton decay matrix elements may strongly depend on the quark mass \cite{Martin:2011nd}. In this work, we use the 2+1 flavor chiral Domain wall fermions at the physical point to reduce systematic uncertainties in the proton decay matrix elements. 

Several next generation experiments are dedicated to search for the proton decay. The Deep Underground Neutrino Experiment (DUNE) is the nearest future experiment to start collecting data. It will use liquid argon time projection chamber, which can track the 3D trajectories of the particles with high precision. DUNE is expected to reach the current bound of proton lifetime quickly, and improve the bound starting from 4 years after the initial run. \cite{Acciarri:2015uup}

We start from the definition of the effective operator for the $|\Delta B| = 1$ baryon number violation process in section~\ref{eff_op}. We calculate the hadronic matrix element of the operator with a proton and a meson final state. In section~\ref{latt}, we describe the lattice ensemble that we use. In section~\ref{result}, we present our results on the proton decay form factor calculation and compare them with the earlier study \cite{Aoki:2017puj}. In section~\ref{con}, we proceed to discuss these results and further calculations we plan to do.

\section{Proton decay matrix elements from the effective operators}
\label{eff_op}
\begin{figure}[h]
\centering \hspace{.5cm}
\includegraphics[width=0.5\linewidth]{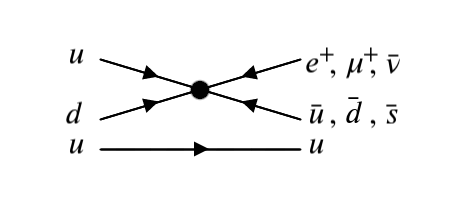}
\caption{Proton decay effective operator}
\label{fig:pdecay}
\end{figure}

(SUSY-) GUT theories can have various types of underlying processes violating the baryon number \cite{Gavela:1981cf,1995phlb..342..138h,sakai:1981pk,grinstein:1982um,babu:1993we,Nath:2006ut,Bajc:2016qcc,Lee:2016wiy}. Regardless of specific theory, one can always express a proton decay rate through matrix elements of the low energy effective operators starting from dimension six as in Fig.~\ref{fig:pdecay}, 
\begin{equation}
    O^I = \epsilon^{ijk} (\mathrm{q}_{i} \mathrm{q}_{j})_{\Gamma}(\mathrm{q}_{k} \ell)_{\Gamma'} ,
\end{equation}
where $\mathrm{q}$ is quark field, $\ell$ is lepton field, and $(\mathrm{q} \mathrm{q})_{\Gamma} = \mathrm{q}^T_{\alpha} (CP_{\Gamma})_{\alpha \beta} \mathrm{q}_{\beta}$ and $C=\gamma_2\gamma_4$ is the charge conjugation matrix in Euclidean spacetime, $P_{\Gamma} = \frac{1 \pm \gamma_5}{2}$ is fermion chirality projection matrix for $\Gamma = L,R$ \cite{Weinberg:1979sa}.
With color and Dirac indices suppressed in the operators, the effective lagrangian is:
\begin{equation}
    \mathcal{L}^{\text{eff}}_{\slashed{B}} =  - \sum_I  C_I O^I
    = \sum_I C_I \bar{\ell}^C \left[ (\mathrm{q}\mathrm{q})_{\Gamma} P_{\Gamma'} \mathrm{q}\right], 
\end{equation}
where $C_I$ is the Wilson coefficient of the operator $O^I$.

We compute the proton decay matrix elements on a lattice between the incoming nucleon state(N) and the outgoing pseudoscalar meson($\Pi = K, \pi$),
\begin{equation}
    \langle \Pi, \ell | O^I | N \rangle = \langle \Pi, \ell | \bar{\ell} (\mathrm{q}\mathrm{q})_{\Gamma} P_{\Gamma'} \mathrm{q} | N \rangle
    = \bar{v}_{\ell} \langle \Pi |  (\mathrm{q}\mathrm{q})_{\Gamma} P_{\Gamma'} \mathrm{q} | N \rangle
\end{equation}
where $\bar{v}_{\ell}$ is the final lepton spinor.
Using the shorthand notation for $O_{\Gamma \Gamma'} = (\mathrm{q}\mathrm{q})_{\Gamma} P_{\Gamma'} \mathrm{q} $ and leaving out the leptonic part, the relevant hadronic parts of the matrix elements for the proton with incoming momentum $p$, the outgoing lepton and meson with momentum $q$ and momentum $p'$ respectively,
\begin{equation}
\langle \Pi(p') | O^{\Gamma \Gamma'}(q) | N(p,s) \rangle = P_{\Gamma '} \left[ W_0^{\Gamma \Gamma'}(q^2) - \frac{i \slashed{q}}{m_N} W_1^{\Gamma \Gamma'} (q^2 )\right] u_N(p,s) 
\end{equation}
can be calculated, from which form factors $W_{0,1}$ can be determined.

With these form factors $W_{0,1}$, the partial decay width of proton $\Gamma$ is given by:
\begin{equation}
\Gamma \left(p \rightarrow \Pi + \bar{\ell} \right) = \frac{(m_p^2 - m_{\Pi}^2)^2}{32\pi m_p ^3}  \left| \sum_I C_I W_0^I\left(p \rightarrow \Pi + \bar{\ell} \right) \right| ^2
\end{equation}
\section{Physical Lattice Ensemble}
\label{latt}
We use the 2+1 flavor $24^3 \times 64 ( L_{\sigma} \sim 4.8 \text{ fm} )$ lattice ensemble generated by RBC/UKQCD collaboration \cite{Mawhinney}, with the Iwasaki-DSDR gauge action and Mobius Domain Wall Fermions with the fifth dimension $L_s = 24$. Lattice cutoff is $a^{-1} = 1.015$ GeV and the quark masses are chosen to be $m_la = 0.00107, m_ha = 0.085$ resulting in the pion mass $m_{\pi} = 0.140$ GeV and kaon mass $m_K = 0.513$ GeV. We employ deflated CG with 2000 compressed eigenvectors using 1000 eigenvectors as a basis \cite{Clark:2017wom}. We use all-modes-averaging (AMA) method in which we approximate the quark propagator by a truncated solution to the Mobius operator \cite{Blum:2012uh,Blum:2012my,Shintani:2014vja}. Per gauge configuration, we compute 1 exact and 32 sloppy samples. With this setup, samples with three source-sink separations $t_{\text{sep}} \in \{ 8,9,10\}$ in lattice units are computed (17 configurations for $t_{\text{sep}}=9$).

The on-shell lepton external states in proton decay require the square of momentum transfer from the initial proton to final meson $-q^2 = m_{\bar{\ell}}^2 $ to be small. While the largest allowed momentum transfer squared takes place at $q^2 = -0.0110 \text{GeV}^2$ when the final lepton is muon, unit momentum on lattice is $0.26$ GeV. Along with energy-momentum conservation, the smallness of $-q^2$ should be achieved by choosing appropriate momenta configuration. Furthermore, to suppress statistical noise, the nucleon sources must carry the smallest momentum possible  \cite{Lepage:1989hf}. We chose momenta configurations such that the kaon momenta are  $\vec{n}_p = (0,0,1),(0,1,1)$, the pion momenta are $\vec{n}_p = (1,1,1), (0,0,2)$, and the proton is at rest, where the momentum on lattice is denoted as $\vec{p} = \frac{2 \pi \vec{n}_p}{L_\sigma } $.

\section{Results}
\label{result}
Matrix elements can be calculated on the lattice using the two point and three point functions defined as follows,
\begin{equation}
\label{threept}
    C_3^{\Gamma \Gamma'}(x',x,x_0) = \langle 0 | J_{\Pi}(t',\vec{x}') O^{\Gamma \Gamma' }(t, \vec{x})  \bar{J}_N (t_0, \vec{x_0} ) |  0 \rangle
\end{equation}
\begin{equation}
    C_{\Pi}(x',x)= \langle 0 | J_{\Pi}(t',\vec{x}')  J^{\dag}_{\Pi}(t, \vec{x}) | 0 \rangle
\end{equation}
\begin{equation}
    C_N (x, x_0) = \langle 0 | J_N(t,\vec{x}) \bar{J}_N(t_0,\vec{x_0})| 0 \rangle .
\end{equation}
The three point correlation function in Eq.~\ref{threept} is constructed by contracting forward propagators with the source at $\vec{x_0}$ and the backward propagator with the sequential source at the meson sink $\vec{x}'$. In this study, we consider only the proton at rest.
To extract the proton decay matrix elements, we use the ratio between the two and three point functions $R^{\Gamma \Gamma'}_3$ :
\begin{equation}
\label{ratiodef}
R_3^{\Gamma \Gamma'} (t',t,t_0; \vec{p'}, q ; P)
=\frac{ \sum_{\vec{x},\vec{x}'} e^{i\vec{p}' \cdot (\vec{x}' - \vec{x_0})} e^{i\vec{q} \cdot (\vec{x} - \vec{x_0})}tr[PC_3^{\Gamma \Gamma'}(x',x,x_0)]}
{ \sum_{\vec{x}',\vec{x}} e^{i\vec{p}' \cdot (\vec{x}' - \vec{x})}C_{\Pi}(x',x) 
\sum_{\vec{x}} e^{i\vec{p} \cdot (\vec{x} - \vec{x_0}) }tr [P_+ C_N(x,x_0)]} \sqrt[]{Z_{\Pi}Z_N} ,
\end{equation}
where $P_+ =\frac{1}{2} (\gamma_4 + 1 ) $, $P$ is a spin projection operator for the proton and $\sqrt{Z_{\Pi}}$, $\sqrt{Z_N}$ are the overlaps of the pseudoscalar and nucleon interpolating operators with their respective states. This ratio approaches the matrix elements 
\begin{equation}
\label{ratio1}
\lim_{t'-t, t-t_0 \rightarrow \infty } 
R_3^{\Gamma \Gamma'} (t',t,t_0; p', q ; P_+ ) =   W_0^{\Gamma \Gamma'}(q^2) 
     - \frac{iq_4}{m_N}W_1^{\Gamma \Gamma'}(q^2)
\end{equation}
\begin{equation}
\label{ratio2}
\lim_{t'-t, t-t_0 \rightarrow \infty } R_3^{\Gamma \Gamma'} (t',t,t_0; p', q ; iP_+\gamma _j ) = 
\frac{ q_{j}}{ m_N} W_1^{\Gamma \Gamma'} (q^2 )
\end{equation}
as the source sink separation is increased.
The relevant form factors $W_0^{\Gamma \Gamma'}$ are computed by taking a linear combination of the ratios with two different projectors $P$. 
\begin{equation}
\label{formfactor}
    W_0^{\Gamma \Gamma'} = R_3^{\Gamma \Gamma'}(t',t,t_0; \vec{p'}, q ; P_+ ) + i\frac{q_4}{q_j} R_3^{\Gamma \Gamma'}(t',t,t_0; \vec{p'}, q ; iP_+\gamma _j )
\end{equation}

\begin{figure}
\begin{subfigure}{7cm}
\includegraphics[width=\linewidth]{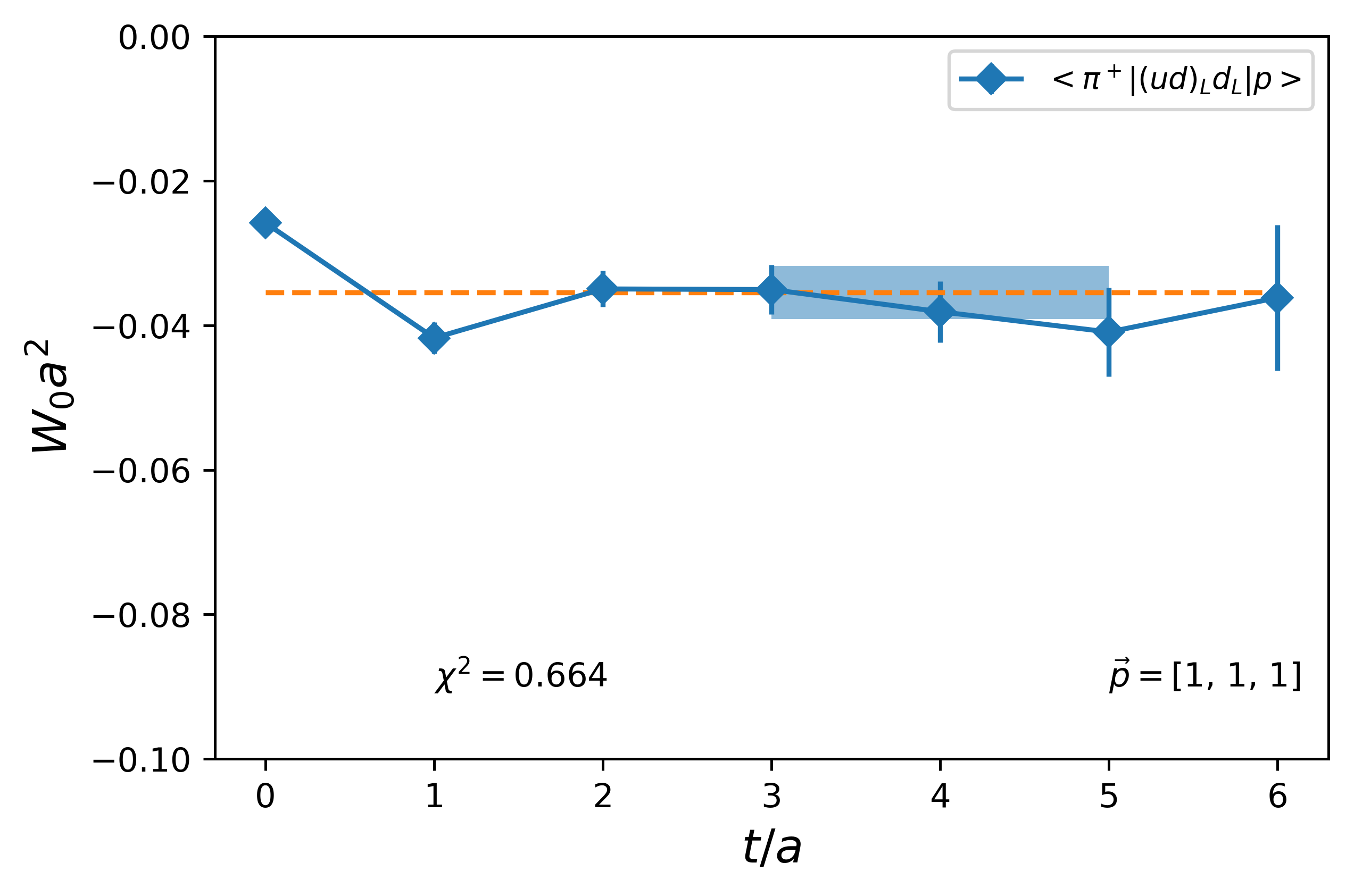}
\caption{ }
\end{subfigure}
\begin{subfigure}{7cm}
\includegraphics[width=\linewidth]{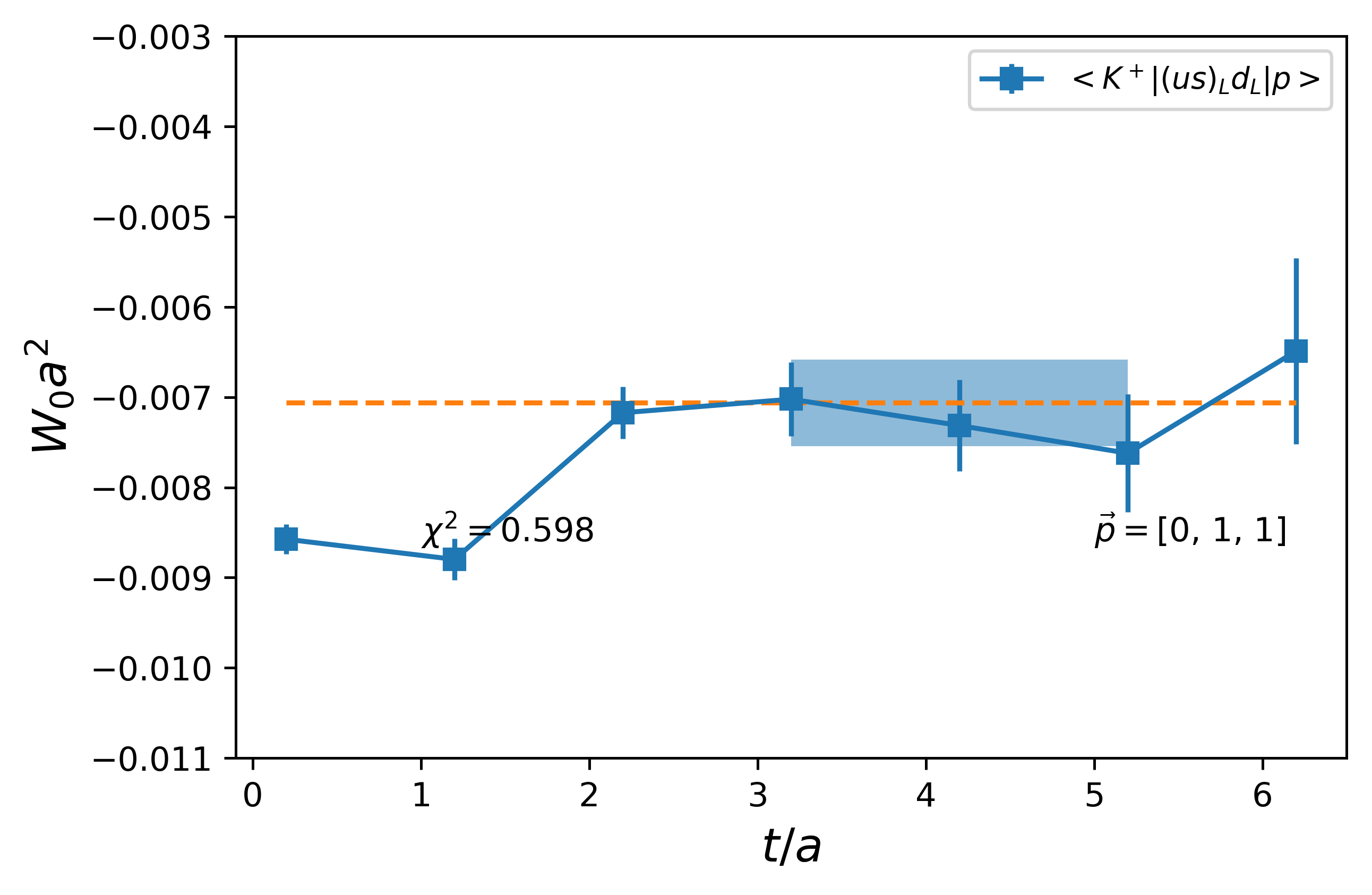}
\caption{ }
\end{subfigure}
\caption{(a) Plateau for the pion channel form factor (b) Plateau for the kaon channel form factor \\
$\vec{p}_{\pi} = [1 1 1]$ $\vec{p}_{K} = [0 1 1]$ $\vec{p}_{p} = [0 0 0]$ at $t_{sep} = 8$ (bare lattice operators)  }
\label{fit}
\end{figure}

\begin{figure}
\begin{subfigure}{0.52\textwidth}
\includegraphics[width=\linewidth]{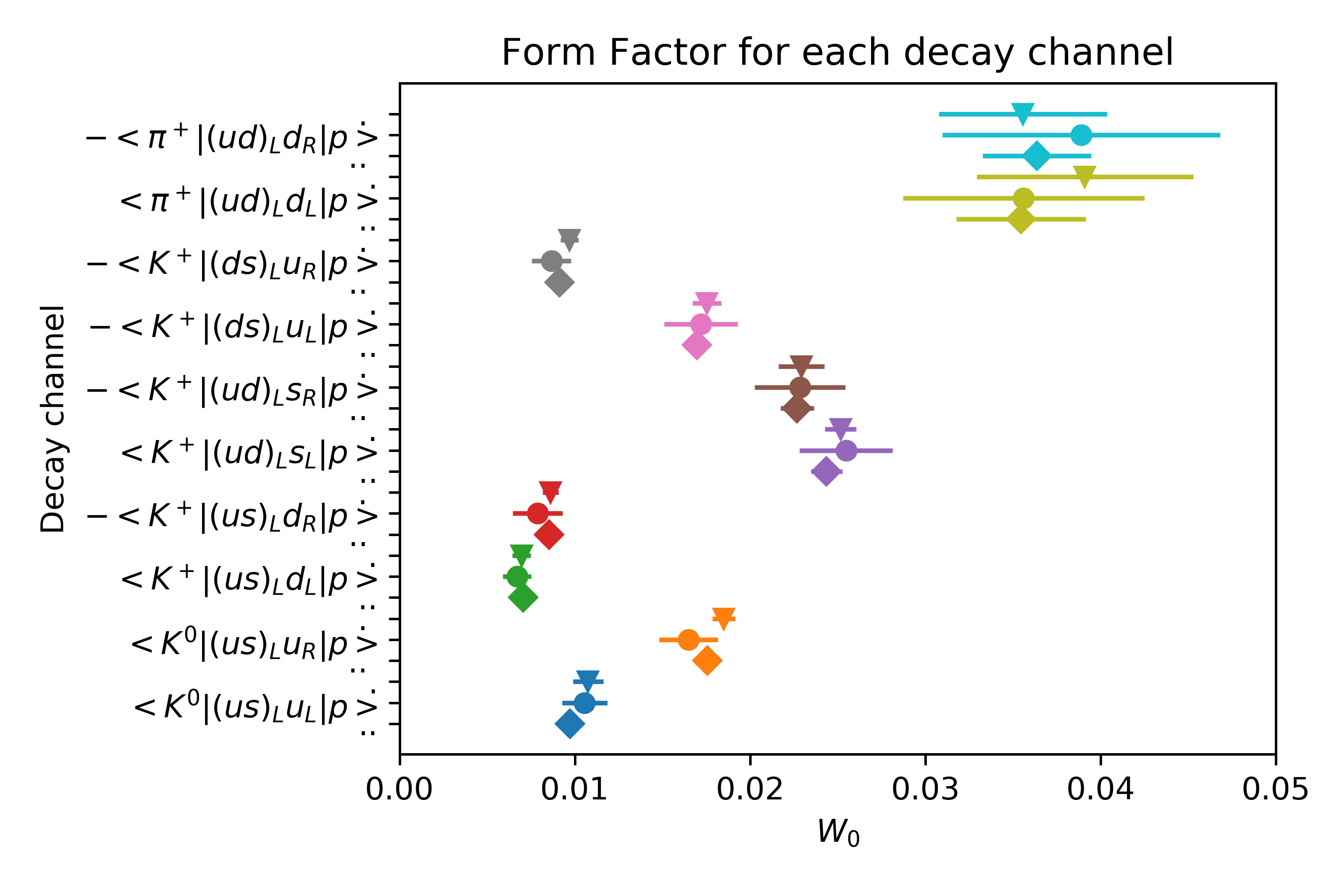}
\caption{ }
\label{channelby}
\end{subfigure}
\begin{subfigure}{0.46\textwidth}
\includegraphics[width=\linewidth]{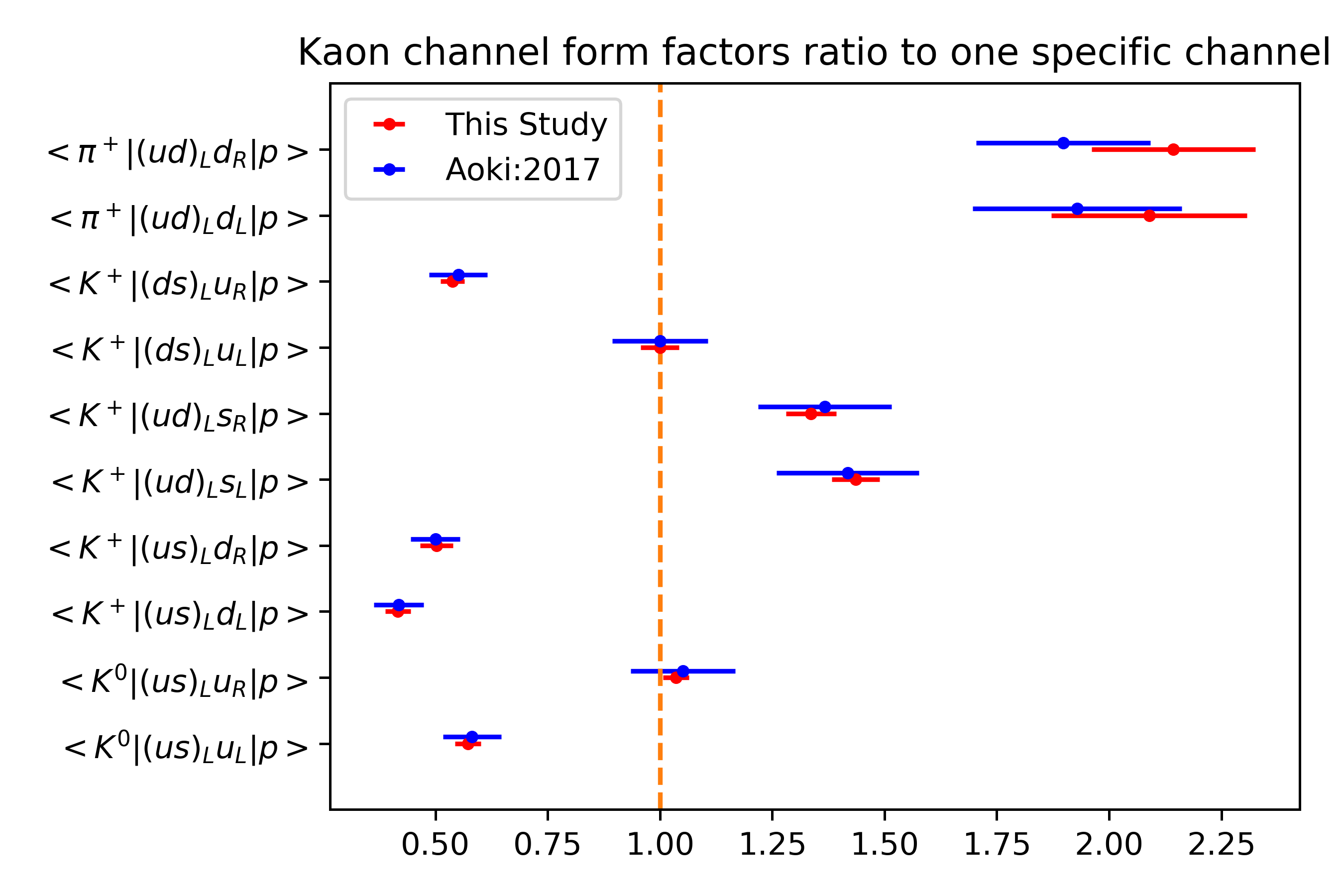}
\vspace{0.18cm}
\caption{ }
\label{Aokicomp}
\end{subfigure}
\caption{ (a) Decay form factors for different channels Symbols represent the source sink separations : $t_{sep} = 8  $ (\protect\markerdiamond), 9(\protect\markercircle), and 10(\protect\invtriangle). (b) Comparison of normalized form factors with our data at $t_{sep} = 8$ and the earlier study \cite{Aoki:2017puj}. }
\label{comparison}
\end{figure}

We calculate the ratio $R_3$ as in (Eq.~\ref{ratiodef}) and $W_0$ as in (Eq.~\ref{formfactor}) to obtain bare lattice values for the proton decay form factors. We fit the matrix elements in Fig.~\ref{fit} with a constant over $t=3$ to $t=5$ with full covariance matrices. Within each Jackknife sample, fit parameters are estimated using covariance matrix on the Jackknife subset. We find that decay form factors $W_0$ for each channel with different source-sink separations are in agreement (see Fig.~\ref{channelby}). Different symbols \markerdiamond, \markercircle, and \invtriangle $ $ mean different source sink separation $t_{sep}= t' - t_0 = $ 8, 9, and 10, respectively.
All the values discussed above are not renormalized.

Since the proton decay matrix elements have only multiplicative renormalization factors \cite{Aoki:1999tw}, we can examine their ratios to compare to the earlier study \cite{Aoki:2017puj} (Fig.~\ref{Aokicomp}). In Ref. \cite{Aoki:2017puj}, the proton decay matrix elements were calculated with chiral 2+1 fermions at heavier pion masses of 340 -- 690 MeV and extrapolated to the physical quark mass. We normalize the proton decay form factors for all channels by $\langle K^+|(ds)_L u_L|p\rangle$ and compare them in Fig.~\ref{Aokicomp}. Errors in Eq.~\ref{eq:aokicomp} are estimated with the propagation of errors in the numerator and the denominator into the errors in the normalized form factors.
\begin{equation}
\label{eq:aokicomp}
W_0^{norm} = \left| \frac{W_0^{\Gamma \Gamma'} (Channel)}{W_0^{\Gamma \Gamma'}
( \langle K^+|(ds)_{\Gamma} u_{\Gamma'}|p\rangle)} \right|
\end{equation}

The error estimates in this preliminary calculation include only statistical uncertainties, while the earlier study included systematic and statistical errors. Each value agrees within errors, indicating the normalized form factors over different channels did not get spoiled by a long chiral extrapolation in Ref~\cite{Aoki:2017puj}. The statistical errors from both studies are compared in columns 2 and 3 of Table 1. For the kaon channels, we have reached a comparable level of statistical errors, while the pion channel shows the errors are larger than in Ref. \cite{Aoki:2017puj}. The systematic errors in the earlier study (columns 3--6 in Table 1)  can be reduced by avoiding the chiral extrapolation. The other large source of error is renormalization, which we will study next.

\begin{table}[ht]
\centering
  \begin{adjustbox}{max width=\textwidth}
  \begin{tabular}{ccc|ccc}
     & \shortstack{Stat. [\%]\\(This study)} & \shortstack{Stat.[\%]\\(Aoki:2017)} & \shortstack{Chiral\\extrapol.[\%] } & $a^2$ [\%] & $\Delta_Z$ [\%]\\
    \midrule
    $\langle K^0|(us)_L u_L|p \rangle$ & 4.97 & 3.5 & 3.1 & 5.0 & 8.1 \\
    $\langle K^0|(us)_L u_R|p \rangle$ & 2.81 & 2.8 & 2.8 & 5.0 & 8.1 \\
    $\langle K^+|(us)_L d_L|p\rangle$ & 6.78  & 4.4 & 7.5 & 5.0 & 8.1\\
    $-\langle K^+|(us)_L d_R|p\rangle$ & 7.28  & 3.7 & 3.5 & 5.0 & 8.1\\
    $\langle K^+|(ud)_L s_L|p\rangle$ & 3.71 & 3.0 & 3.9 & 5.0 & 8.1\\
    $-\langle K^+|(ud)_L s_R|p\rangle$ & 4.21 & 3.2 & 1.6 & 5.0 & 8.1\\
    $-\langle K^+|(ds)_L u_L|p\rangle$ & 4.25 & 2.8 & 2.1 & 5.0 & 8.1\\
    $-\langle K^+|(ds)_L u_R|p\rangle$ & 4.98 & 3.6 & 2.7 & 5.0 & 8.1\\
    $\langle \pi^+|(ud)_L d_L|p\rangle$ & 10.44 &  3.4 & 5.7 & 5.0 & 8.1\\
    $-\langle \pi^+|(ud)_L d_R|p\rangle$ & 8.52 &  3.0 & 1.8 & 5.0 & 8.1\\
  \end{tabular}
  \end{adjustbox}
  \caption{ Left : Comparison of statistical errors. Right: Systematic errors in chiral extrapolation, $O(a^2)$, $\Delta_Z$ ( \cite{Aoki:2017puj}) }
\end{table}

\section{Discussion and outlook}
\label{con}
Our preliminary calculation shows that the proton decay matrix elements can be computed on a lattice at the physical point with good statistical precision.
We find that the proton decay matrix elements computed with three different source-sink separations are in agreement indicating negligible excited state effects. The ratios of the proton decay matrix elements for different channels also agree with those from the earlier calculation. The strong dependence suggested in \cite{Martin:2011nd} on the quark mass in the proton decay matrix elements does not appear in the normalized form factors. However, these are only preliminary results of our simulation and they require additional steps to have solid physical meaning. Increasing the number of samples will reduce the statistical errors. Multi-state fits will be done to control the excited states contamination. Non-perturbative renormalization will be computed to obtain values in a continuum renormalization scheme.

We plan to explore other channels to which DUNE will be sensitive, such as decay of proton into two mesons and a lepton. These additional channels can be easily implemented on a lattice, however, studying two-meson final states will require recently developed lattice methodology \cite{Briceno:2014uqa}. Studying the three body channel of proton decay will be relevant for the future experiments. 
The rate of the three body decays $p \rightarrow \pi \pi e^+
$ have been estimated using the effective pion-nucleon interaction $ \mathcal{L}^{\text{eff}} = g_r (\bar{N}\vec{\tau} \cdot \vec{\pi} N) $ and may be 140\% of the rate of the two body decay for isospin 0 state and 24\% for isospin 1 state \cite{Wise:1980ch}. The SUSY-GUT model estimation of the $p \rightarrow K \pi \ell $ three body decay predicts the decay width to be 7 --15 \% of the two body decay width \cite{Vissani:1995hp}.

\acknowledgments We are grateful to RBC/UKQCD collaboration for QCD ensemble,  discussions and supports. 
SS also acknowledges support by the RHIC Physics Fellow Program of the RIKEN BNL Research Center.
This computational resources in this work are provided by the Scientific Data and Computing Center (SDCC),  Computational Science Initiative (CSI) at BNL, and Jefferson Laboratory through USQCD.  This work is supported in part by US DOE Contract \#AC-02-98CH10886(BNL) and  JSPS KAKENHI Grant \#16K05320 and \#17H02906. 

\bibliographystyle{aip}
\bibliography{ref}

\end{document}